\documentclass[aps,twocolumn,superscriptaddress,floatfix,longbibliography]{revtex4-1}

\usepackage{bm,amsmath,amsfonts,amssymb,braket}
\usepackage{latexsym,dsfont,array,layout,mathrsfs,textcase}
\usepackage{graphicx,graphics,subfigure,xcolor}
\usepackage{comment,verbatim}
\usepackage{float}
\usepackage{times}
\usepackage{xspace}
\usepackage{stmaryrd}
\usepackage{multirow}
\usepackage{mathtools}
\usepackage{booktabs}
\usepackage{CJK}
\usepackage[colorlinks,linkcolor=blue,citecolor=blue,urlcolor=blue]{hyperref}

\begin{document}

\title{Phase induced localization transition}
\begin{CJK*}{UTF8}{gbsn} 

\author{Tong Liu ({\CJKfamily{gbsn}刘通})} %
\affiliation{School of Science, Nanjing University of Posts and Telecommunications, Nanjing 210003, China}%

\author{Xingbo Wei ({\CJKfamily{gbsn}魏兴波})}%
\thanks{weixingbo@zstu.edu.cn}
\affiliation{Department of Physics and Key Laboratory of Optical Field Manipulation of Zhejiang Province, Zhejiang Sci-Tech University, Hangzhou 310018, China}%

\author{Youguo Wang ({\CJKfamily{gbsn}王友国})}%
\thanks{wangyg@njupt.edu.cn}
\affiliation{School of Science, Nanjing University of Posts and Telecommunications, Nanjing 210003, China}%

\date{\today}

\begin{abstract}
Localization phenomenon is an important research field in condensed matter physics. However, due to the complexity and subtlety of disordered syestems, new localization phenomena always emerge unexpectedly. For example, it is generally believed that the phase of the hopping term does not affect the localization properties of the system, so the calculation of the phase is often ignored in the study of localization. Here, we introduce a quasiperiodic model and demonstrate that the phase change of the hopping term can significantly alter the localization properties of the system through detailed numerical simulations such as the inverse participation ratio and multifractal analysis. This phase-induced localization transition provides valuable information for the study of localization physics.

\vspace{0.5cm}
\noindent $ \textbf{Keywords:} $ phase, localization, quasiperiodic

\vspace{0.5cm}
\noindent $ \textbf{PACS:} $ 71.23.Ft, 71.10.Fd, 71.23.An

\end{abstract}

\maketitle
\end{CJK*}

\section{Introduction}

The concepts of topology and localization in condensed matter physics are pivotal areas of study, which reveal profound insights into the behavior of quantum systems~\cite{topo3,topo4,Anderson}. 
Topological phenomena~\cite{topo5} in materials are characterized by global properties that remain unchanged under continuous deformations, providing a robust framework for understanding quantum states that are resistant to local perturbations.
Localization~\cite{disorder1}, on the other hand, refers to the phenomenon in which disorder within a system leads to spatial confinement of wave functions, significantly impacting the transport characteristics of electrons.
The interaction between topology and localization not only enhances our theoretical comprehension, but also lays the groundwork for practical applications in quantum computing and materials science. This interplay~\cite{tp_AAH_1,tp_AAH_2} is particularly fascinating as it is essential for the development of advanced technological devices.
And there exist numerous exquisite toy models in the field, with the Harper-Hofstadter model~\cite{Hofstadter} and Aubry-Andr\'{e}-Harper model~\cite{AAH1, AAH2} serving as archetypes of low-dimensional topological physics and localized physics, respectively.

The Harper-Hofstadter (HH) model, also known as the Hofstadter butterfly model, investigates the quantum behavior of electrons in a two-dimensional lattice subjected to a perpendicular magnetic field. The model~\cite{Hofstadter}, initially proposed by Douglas Hofstadter in 1976, offers valuable insights into the behavior of electrons in magnetic fields and carries significant implications for the study of topological phases of matter and quantum computing~\cite{Hofstadter1}. 
In the HH model, electrons traverse a square lattice in which each lattice site represents an atom, allowing for electron hopping between sites. The presence of a perpendicular magnetic field introduces a phase factor to the hopping terms due to the magnetic vector potential, which is proportional to the magnetic flux through the lattice. The most remarkable characteristic of the HH model is its fractal energy spectrum, famously depicted as the Hofstadter butterfly.~\cite{Hofstadter2}.
\begin{figure*}
  \centering
  \includegraphics[width=1\textwidth]{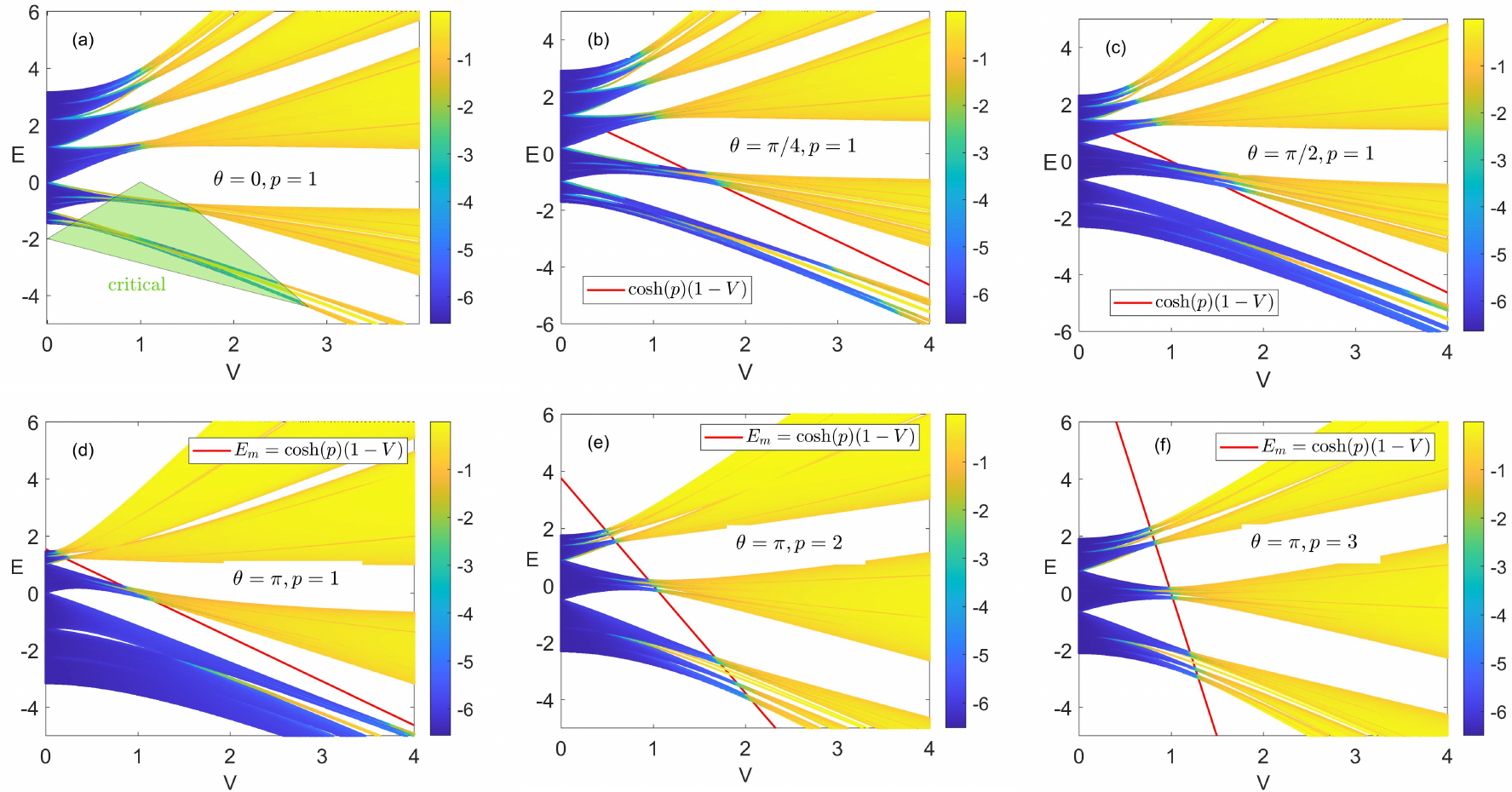}\\
  \caption{(Color online) The eigenvalues and $\log(\rm IPR)$ of Eq.~(\ref{eq1}) as a function of $V$. Different colors of the eigenvalue curves indicate different magnitudes of $\log(\rm IPR)$. The blue region denotes extended states, and the yellow region denotes localized states. (a) When $\theta=0$ and $p=1$, the system has a large number of critical states, as seen in the green region, which is completely consistent with Ref.~\cite{critical}. (b) and (c)
   When $0<\theta<\pi$ and $p=1$, there exists a delocalization-localization phase transition in the energy spectrum, but the phase boundary is not clearly defined. (d) When $\theta=\pi$ and $p=1$, the delocalization-localization phase boundary in the energy spectrum becomes clear and regular, the red solid line represent the estimated mobility edge $E_m=\cosh(p)(1-V)$. (e) and (f) illustrate that the estimated mobility edge is applicable to other $p$-values when $\theta=\pi$.
   The total number of sites is set to be $L=987$. }
  \label{fig1}
\end{figure*}

The Aubry-Andr\'{e}-Harper model (AAH) model studies the localization phenomena in quasiperiodic systems~\cite{liu1,liu2}.  
The model~\cite{AAH1,AAH2} introduced by S. Aubry and G. Andr\'{e} in 1980 explores the transition between extended and localized states in a one-dimensional lattices with a quasiperiodic potential, which is intermediate between completely periodic and disordered systems.
The self-duality of the AAH model is a significant aspect, as it maintains invariance under a Fourier transform, exchanging the roles of position and momentum in the mathematical form of the Hamiltonian~\cite{liu4,cai}. 
The AAH model not only enhances our comprehension of quasiperiodic systems, but also acts as a link between the physics of ordered and disordered systems, providing a fertile area for exploration in both theoretical and experimental condensed matter physics~\cite{wei1,wei2}.

The primary connection between the HH and AAH models lies in their shared emphasis on the impact of competing periodicities~\cite{QHI1}. The HH model explores the interaction between lattice periodicity and magnetic field periodicity, while the AAH model investigates the effects of quasiperiodicity.
Both models provide valuable insights into phenomena such as Anderson localization and the quantum Hall effect~\cite{QHI2}, illustrating the ways in which different scales and types of periodicity can impact electronic properties in disordered and externally magnetized systems.
The hopping terms in the AAH model are characterized by zero phase factors, while the HH model can be effectively reduced to a one-dimensional AAH-like model with non-zero phase factors. The presence of non-zero phase factors significantly influences the topological properties of the HH model. Contrastively, it is widely accepted that the phase factor has no impact on the localization properties of the AAH model or quasiperiodic systems~\cite{tp_AAH_3,tp_AAH_4}. A natural question arises: do phase factors have an impact on localization? In this study, we introduce a quasiperiodic system and demonstrate that the phase of the hopping terms can also significantly alter the localization properties of the system.

\section{Model and results}

Consider an exponentially decaying hopping model with a phase factor, characterized by the discrete Schr\"{o}dinger equation
\begin{equation}
\begin{aligned}
&\sum _{l=1}^{\infty } e^{p(1- l)}( e^{-i \theta}\psi_{n-l}+e^{i \theta}\psi_{n+l}) + \\
& 2 V\sum _{l=1}^{\infty }  e^{p(1- l)} \cos (2 \pi  \alpha l n) \psi_n=E \psi_n,\\ 
\label{eq1}
\end{aligned}
\end{equation}
where $p>0$ represents the exponential decay parameter, $V$ is the quasiperiodic potential strength, and $\psi_n$ is the amplitude of wave function at the $n$th lattice, $l$ takes positive integers. We choose the irrational number $\alpha=(\sqrt{5}-1)/2$.
When the phase $\theta$ is set to 0, Eq.~(\ref{eq1}) becomes the quasiperiodic model studied in Ref.~\cite{critical}, which hosts a generalized duality symmetry and rich critical states. In this work, we focus on the $\theta \neq 0$ case.

To acquire the eigenvalue $E$ and the associated wave functions $\psi$, we firstly numerically diagonalize
the Schr\"{o}dinger equation [Eq.~(\ref{eq1})]. Then, it is convenient to calculate physical quantities through comprehensive numerical analysis.
The localization of a wavefunction $\psi$ can be determined by the inverse participation ratios~\cite{liu3,wei3,cai1},
\begin{equation}
{\rm IPR}_{j}=\sum_{n=1}^{L}|\psi^{(j)}_{n}|^4,
\end{equation}
where $\psi$ has been normalized, $L$ represents the total number of lattice sites, and $j$ denotes the energy level index.
Localized states are characterized by an IPR that is independent of the system size $L$ due to their unaffected nature by boundaries. In contrast, extended states exhibit a scaling with the system size, leading to corresponding IPRs that scale like $L^{-1}$ and tend towards 0. While critical states demonstrate multifractality and are characterized by the scaling exponent between 0 and 1.

To highlight the distinction between extended, critical, and localized states, we analyze the IPR scaling exponent, $\log(\rm IPR)$, instead of focusing solely on the raw value of the IPR.
Figure \ref{fig1} depicts the spectrum $E$ of various parameters $\theta$ and $p$ as a function of the quasiperiodic potential strength $V$, with consideration given to a system size of $L=987$.
The color representation indicates the $\log(\rm IPR)$ of the corresponding eigenvalues, specifically, the blue region signifies that the corresponding eigenstates are extended. On the contrary, the yellow region indicates that the corresponding eigenstates are localized. As depicted in Fig.~\ref{fig1} (a), the spectrum of Eq.~(\ref{eq1}) [$\theta=0$ and $p=1$] is composed of multiple bands, for $E>0$, the extended-localized transition point occurs at $V=1$, while for $E<0$, the transition point transforms into a critical region (green area). All of these numerical findings align well with the predictions in Ref.~\cite{critical}.
\begin{figure}
  \centering
  \includegraphics[width=0.48\textwidth]{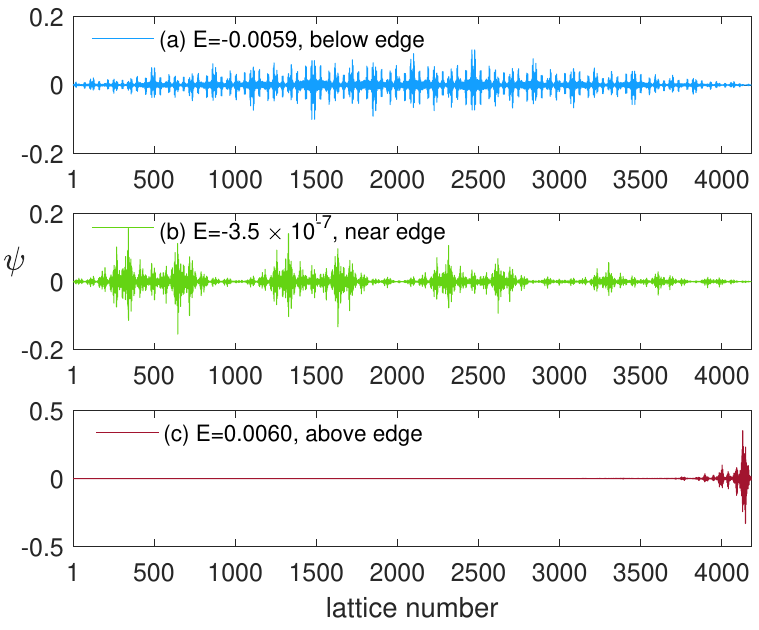}\\
  \caption{(Color online) Eigenstates numerically obtained from Eq.~(\ref{eq1}) [$\theta=\pi$] with $p=1$, $V=1$ and $L=4181$, and the estimated mobility edge $E_{m}=0$. Three different eigenvalues: (a) low energy extended state below  $E_{m}=0$, (b) critical state near $E_{m}=0$, and (c) high energy localized state above $E_{m}=0$.
  }
  \label{fig2}
\end{figure}
\begin{figure*}
  \centering
     \subfigure{
    \begin{minipage}[]{0.48\textwidth}
    \centering
    \includegraphics[width=1\textwidth]{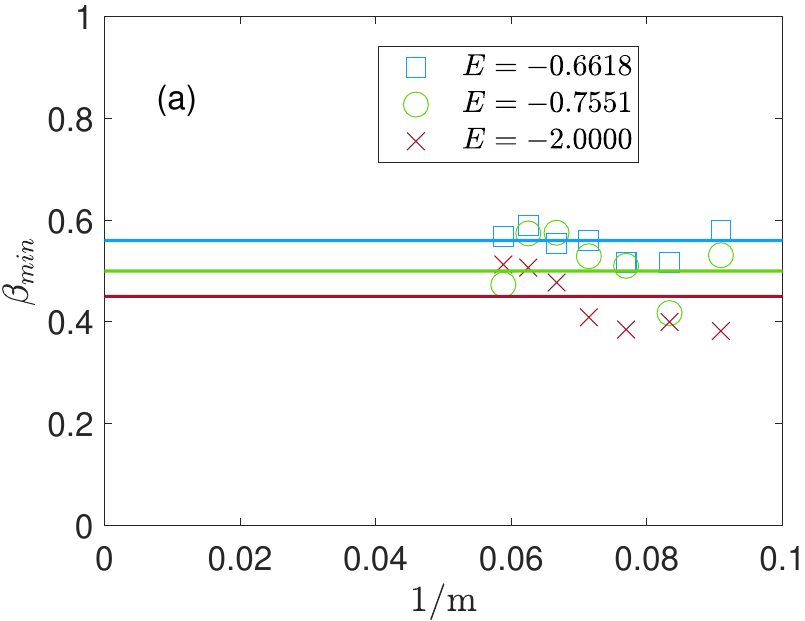}
    \end{minipage}}
  \subfigure{
    \begin{minipage}[]{0.48\textwidth}
    \centering
    \includegraphics[width=1\textwidth]{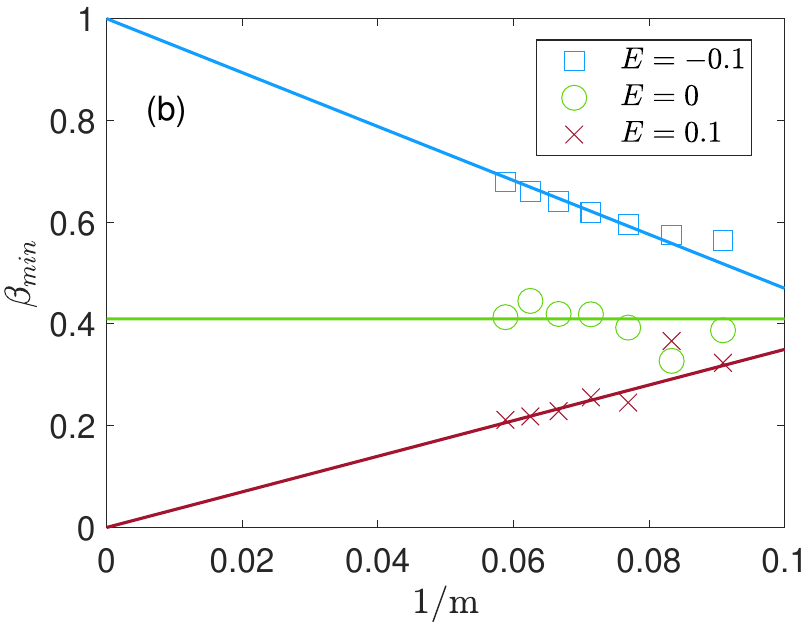}
    \end{minipage}}
  \caption{(Color online) (a) $\beta_{min}$ as a function of the inverse Fibonacci index $1/m$ for the $\theta=0$ case. When $V=1$ and $E<0$, the eigenstates are critical states. (b) $\beta_{min}$ as a function of the inverse Fibonacci index $1/m$ for the $\theta=\pi$ case. The green markers correspond to the critical state ($E=0$), the blue markers correspond to the extended state ($E=-0.1$), and the red markers correspond to the localized state ($E=0.1$).
  }
  \label{fig3}
\end{figure*}

When $0<\theta<\pi$ and $p=1$, the spectrum of Eq.~(\ref{eq1}) undergoes a dramatic change, a irregular delocalization-localization phase transition occurs in the energy spectrum, as shown in Fig.~\ref{fig1} (b) and (c). 
Surprisingly, when $\theta=\pi$ and $p=1$, there is no extended-localized transition observed around $V=1$, and the critical regions also disappear. 
Instead, a regular mobility edge emerges to delineate the boundary between extended and localized states, with the estimated expression for the mobility edge being $E_m=\cosh(p)(1-V)$, as shown in Fig.~\ref{fig1} (d). To verify the robustness of the mobility edge, we conduct numerical simulations for various $p$-values and demonstrate that the estimated expression also fits very well, thus proving its robustness, as shown in Fig.~\ref{fig1} (e) and (f).
We have also explored alternative parameter combinations and found that except for $\theta=0$ and $\pi$, all other values of $\theta$ have a corresponding pair with $\theta+\pi$, such as the localization properties of $\theta=\pi/2$ being the same as those of $\theta=3\pi/2$. These numerical findings indicate that substantial changes in the system's localization occur as the phase of the hopping terms varies, a novel phenomenon not previously observed in studies on disordered systems.

To further confirm the presence of the mobility edge, we illustrate the spatial distributions of related eigenstates with eigenvalues close to $E_m$.
We specify the parameters $p=1$, $V=1$, and $L=4181$ for performing the numerical calculation. We select three typical wave functions with corresponding eigenvalues that are above, below, and near the mobility edge $E_{m}=0$. 
Figure~\ref{fig2} illustrates the spatial distributions of different eigenstates. As depicted in Fig.~\ref{fig2}(a), the eigenstate exhibits intuitive extension, with the corresponding eigenvalue below $E_{m}=0$. In Fig.~\ref{fig2}(c), the eigenstate corresponds to localized states, with an eigenvalue above $E_{m}=0$. Notably, in Fig.~\ref{fig2}(b), the eigenvalue is close to $E_{m}=0$, and the eigenstate clearly demonstrates self-similar properties, indicating that this state is neither fully localized nor extended over the entire space, namely it is a critical state.

\section{Criticality and multifractal analysis}
The eigenstates at the mobility edge in the case of $\theta=\pi$ exhibit critical behavior, while in the case of $\theta=0$, there is also a significant number of critical states in the region where $E<0$.
To better compare the localization properties of these two cases, we employ multifractal analysis~\cite{fractal_1,fractal_2} to calculate the eigenstate properties. 
Multifractal analysis expands upon the concept of fractals, which are geometric structures displaying self-similar patterns across different scales. This approach is particularly valuable in quasiperiodic systems, where scale-invariant properties are manifested.

For each wave function $\psi_n^{(j)}$, a scaling exponent $\beta_{n}^{(j)}$ can be derived from the $n$th probability $P_{n}^{(j)} = \vert\psi_n^{(j)} \vert^2 \sim (1/F_{m})^{\beta_{n}^{(j)}}$. According to multifractal analysis, in the case of an extended eigenstate, the maximum of $P_{n}^{(j)}$ scales as $\max(P_{n}^{(j)})\sim (1/F_{m})^1$, indicating that $\min(\beta_{n}^{(j)})=1$. Conversely, for a localized eigenstate, $P_{n}^{(j)}$ peaks at a single lattice and is close to zero at other lattices, resulting in $\max(P_{n}^{(j)}) \sim (1/F_m)^0$, or equivalently $\min(\beta_{n}^{(j)})=0$. As for critical eigenstates, the corresponding value of $\beta_{min}^{(j)}$ falls within the interval $(0,~1)$ and can serve as a distinguishing factor between extended and localized states.

Figure~\ref{fig3} illustrates the dependence of $\beta_{min}$ on the inverse Fibonacci index $1/m$. In Fig.~\ref{fig3}(a), it is evident that for the case of $\theta=0$, $V=1$ and $E<0$, $\beta_{min}$ lies within the range $(0, 1)$ in the limit of large $L$, consistent with the findings in Fig.~\ref{fig1}(a). As shown in Fig.~\ref{fig3}(b), when $\theta=\pi$ and $E=0$, $\beta_{min}$ also remains within $(0, 1)$ for large $L$, indicating critical behavior near the mobility edge $E_m=0$. For an eigenvalue of $E=-0.1$, $\beta_{min}$ approaches 1 as $L$ becomes large, suggesting extended states. Conversely, for an eigenvalue of $E=0.1$, $\beta_{min}$ asymptotically tends to 0 as $L$ increases, indicating localized states. We have also checked other combinations of parameters and get the same results as expected.

The above numerical results illustrate that altering the phase of the hopping term significantly impacts the localization properties of the system. Specifically, a critical region ($\theta=0$) containing numerous critical states is transformed into a region with a mobility edge ($\theta=\pi$). This phenomenon can be referred to phase induced localization transition.

\section{Summary}
In summary, we present a quasiperiodic model incorporating the hopping phase, and thoroughly analyze the localization properties of this model. Our findings reveal that when the phase is set to 0, the system displays numerous critical states; conversely, setting the phase to $\pi$ results in the appearance of a mobility edge that separates extended and localized states in the spectrum. This highlights the significant role of an initially overlooked phase in this system, leading to a remarkable change in its localization properties. Our findings reveal a new phenomenon in localized physics and lay the groundwork for further elucidating the underlying physical mechanisms.

\begin{acknowledgments}
This work was supported by the National Natural Science Foundation of China (Grant No. 62071248), the Zhejiang Provincial Natural Science Foundation of China
(Grant No. LQ24A040004), Natural Science Foundation of Nanjing University of Posts and Telecommunications (Grants No. NY223109), and China Postdoctoral Science Foundation (Grant No. 2022M721693).
\end{acknowledgments}


\end{document}